\documentclass[conference]{IEEEtran}

\usepackage[pdftex]{graphicx}
\usepackage{footnote}

\usepackage{algorithm}
\usepackage{algorithmicx}
\usepackage[noend]{algpseudocode}
\algnewcommand{\LeftComment}[1]{ \(\triangleright\) #1}

\makesavenoteenv{tabular}
\makesavenoteenv{table}

\usepackage[caption=false,font=normalsize,labelfont=sf,textfont=sf]{subfig}

\usepackage{float}

\usepackage{url}

\usepackage{amsmath}
\newcommand{\passport}{\textsc{DEFenD}}

\usepackage{array}
\newcolumntype{L}[1]{>{\raggedright\let\newline\\\arraybackslash\hspace{0pt}}m{#1}}
\newcolumntype{C}[1]{>{\centering\arraybackslash}p{#1}}

\begin{document}

\title{\passport{}: A Secure and Privacy-Preserving\\Decentralized System for Freight Declaration}

\author{
	\IEEEauthorblockN{
   	D. Vos,
       L. Overweel,
       W. Raateland,
       J. Vos,
       M. Bijman,
       M. Pigmans, 
       Z. Erkin
	}
    
   \IEEEauthorblockA{
       \\Cyber Security Group, Department of Intelligent Systems\\Delft University of Technology, The Netherlands
	}
    
   \IEEEauthorblockA{
   	\\\{D.A.Vos, L.P.Overweel, W.Raateland, J.V.Vos, M.D.Bijman, M.Pigmans\}@student.tudelft.nl, Z.Erkin@tudelft.nl
   }
}

\maketitle

\begin{abstract}

Millions of shipping containers filled with goods move around the world every day. Before such a container may enter a trade bloc, the customs agency of the goods' destination country must ensure that it does not contain illegal or mislabeled goods. Due to the high volume of containers, customs agencies make a selection of containers to audit through a risk analysis procedure. Customs agencies perform risk analysis using data sourced from a centralized system that is potentially vulnerable to manipulation and malpractice. Therefore we propose an alternative: \passport{}, a decentralized system that stores data about goods and containers in a secure and privacy-preserving manner. In our system, economic operators make claims to the network about goods they insert into or remove from containers, and encrypt these claims so that they can only be read by the destination country's customs agency. Economic operators also make unencrypted claims about containers with which they interact. Unencrypted claims can be validated by the entire network of customs agencies. Our key contribution is a data partitioning scheme and several protocols that enable such a system to utilize blockchain and its powerful validation principle, while also preserving the privacy of the involved economic operators. Using our protocol, customs agencies can improve their risk analysis and economic operators can get through customs with less delay. We also present a reference implementation built with Hyperledger Fabric and analyze to what extent our implementation meets the requirements in terms of privacy-preservation, security, scalability, and decentralization.

\end{abstract}

\IEEEpeerreviewmaketitle

\section{Introduction}

In today's economy, many countries have specialized in producing certain goods: the Netherlands grows the most tulips, China assembles iPhones, and Honduras is the biggest producer of coffee \cite{cia-world-factbook}. These exports are consumed by people all over the world, so millions of containers full of goods move in and out of the world's ports every day \cite{UN16}.

Before any goods may enter a trade bloc, they must be cleared by the relevant customs agency. The customs agency at the destination country of the goods taxes the goods, and attempts to prevent forbidden goods from entering their trade bloc. However, customs agencies are only able to audit on the order of 1\% of incoming containers due to the high volume of containers they are processing \cite{wiebes17}. Therefore, customs agencies must determine which small portion of the containers to examine.

To decide which containers to examine, the customs agencies estimate for each container the risk that it is carrying illegal or mislabeled goods. Customs agencies audit those containers with the highest estimated risk. This risk analysis depends heavily on the available data and its quality and reliability.

In a typical scenario, there are two types of parties involved. The \emph{economic operators} move goods in and out of containers and move containers around the world. The \emph{customs agencies} need reliable data on these goods and container movements for their risk analysis calculations. The economic operators create data about their goods and containers, and customs agencies consume that data. The transfer of data from economic operator to customs agency in the current system is based around a \emph{bill of lading}.

A bill of lading is an aggregate of information about all the goods on a single shipment of containers coming into a port. It is created by the economic operator in charge of that shipment, and sent to the relevant customs agency at least 24 hours before the ship arrives in the port. This system has three major shortcomings:

\begin{itemize}
	\item The bill of lading does not tell a customs agency through which other ports a container of goods may have traveled.
    \item The bill of lading is an aggregation, so it is not the original source of data.
    \item The bill of lading is only required to be received 24 hours before a ship's arrival.
\end{itemize}

These shortcomings make it more difficult for customs agencies to predict which containers must be audited. \\

\noindent A naive solution to this problem would be to create a central trusted authority that collects data from all economic operators from the various online locations where it is available \cite{dimitrova13survey}. Such a centralized system exist: ConTraffic, a  ``web-based geographical information system enabling interactive visualization of container movements'' collects data by mining public data repositories of economic operators \cite{dimitrova13}. This centralized approach has several drawbacks. A central authority could alter the data, and could decide to exclude or mistreat specific economic operators. Therefore, such a centralized system requires \emph{trust}, which cannot be expected of all economic operators and customs agencies in the world.

These security concerns raise the need for a decentralized system, in which involved parties put their trust into a system rather than an organization. This system should enable different, mutually untrusted entities to collaborate and be privacy-preserving, secure, scalable, and decentralized. \\

\begin{table*}
  \renewcommand{\arraystretch}{1.3}
  \centering
  \caption{Comparison of blockchain implementations.}
  \label{table:blockchain-comparison}
  \begin{tabular}{ p{2.5cm} p{3.5cm} l l l }
    \hline
    \textbf{Name} & \textbf{Maintainer} & \textbf{Permission} & \textbf{Consensus} & \textbf{GitHub} \\
    \hline
    Chain & Chain Inc & Permissioned & Federated Consensus \cite{consensus-fc} & \url{chain/chain} \\
    Corda & R3 & Permissioned & (Custom) & \url{corda/corda} \\
    Ethereum & Ethereum Foundation & Both & Proof-of-Work \cite{nakamoto08} & \url{ethereum/go-ethereum} \\
    Hyperledger Fabric & The Linux Foundation, IBM & Permissioned & (Custom) & \url{hyperledger/fabric} \\
    Hyperledger Iroha & The Linux Foundation & Permissioned & Byzantine Fault Tolerance \cite{Vukolic16} & \url{hyperledger/iroha} \\
    Hyperledger Sawtooth lake & The Linux Foundation, Intel Corporation & Both & Proof of Elapsed Time \cite{consensus-poet} & \url{hyperledger/sawtooth-core} \\
    Kadena & Kadena LLC & Permissioned & ScalableBFT \cite{consensus-sbft} & (closed source) \\
    MultiChain & Coin Sciences Ltd & Permissioned & Practical BFT \cite{consensus-pbft} & \url{multichain/multichain} \\
    OpenChain & Coinprism & Permissioned & Partionned Consensus \cite{consensus-pc} & \url{openchain/openchain} \\
    Quorum & JPMorgan Chase \& Co. & Permissioned & Raft \cite{consensus-raft} & \url{jpmorganchase/quorum} \\
    Ripple & Ripple & Permissioned & Ripple \cite{consensus-consensus} & \url{ripple/rippled} \\
    Tendermint & All In Bits, Inc. & Permissioned & Byzantine Fault Tolerance \cite{Vukolic16} & \url{tendermint/tendermint} \\
    \hline
  \end{tabular}
\end{table*}

\noindent We propose a decentralized system named \passport{}: \emph{a secure and privacy-preserving \textsc{DE}centralized system for \textsc{F}reight \textsc{D}eclaration}, which enables economic operators and customs agencies to collaborate in an environment that does not require centralized trust, also when they do not have a direct business connection.

In our proposal, economic operators share data about containers and the goods within them through the network. When an economic operator inserts or removes goods from a container, they send this information to the network as an encrypted \emph{claim}. This claim can only be decrypted by the customs agency at the goods' country of destination. Whenever a container changes hands, the involved economic operators create a signed unencrypted claim specifying the involved parties, as well as when and where this happened. Any customs agency can then observe the whole history of the container in question. 

While the entire network can observe the container movements, the package information is only available to the destination agency. This preserves the privacy of the economic operators. Any alterations or mismatching data about the container movements can then easily be detected, resulting in a significant increase in the detection of high-risk containers.

In this paper, to the best of our knowledge, we propose the first decentralized system for freight declaration. The mechanism currently in use is not reliable, causing significant loss in container fraud detection accuracy. Our proposal presents a secure, privacy-aware, scalable system that solves the fundamental trust problem in the container shipment industry. We present the design details of the system using available blockchain technology, and introduce the key data partitioning schema that allows for validation while preserving key privacy requirements. We believe our proposal will enable a number of customs agencies and economic operators to start collaborating in freight declaration without having to trust any one authority to keep their data secure.

The rest of the paper is organized as follows. In Section \ref{section:background}, we present the building blocks of our system; in Section \ref{section:protocol}, we describe our proposal; in Section \ref{section:implementation} we detail and analyze our system design; and finally, we conclude in Section \ref{section:conclusion}.

\section{Background}
\label{section:background}


A \emph{blockchain} is often referred to as a ``shared ledger'', which is a type of distributed database technology. Many \emph{nodes} form a peer-to-peer network that maintains this \emph{shared ledger} consisting of \emph{transactions}. These nodes use a \emph{consensus algorithm} to determine which data may be added to their copy of the shared ledger.

\emph{Nodes} in a blockchain network are responsible for maintaining the data in the shared ledger. In a \emph{permissionless} blockchain, such as Bitcoin \cite{nakamoto08}, any internet-connected computer, capable of understanding the blockchain network's protocol can participate in the network. In a \emph{permissioned} blockchain, only select nodes may participate.

A \textit{shared ledger} is a distributed append-only database present on each node. Data is added to the ledger in the form of \emph{blocks}. As a new block is added to the ledger, the nodes synchronize their copies of the shared ledger by applying the consensus algorithm. In a \emph{public} blockchain, anyone can read the data in the shared ledger, while in a \emph{private} blockchain, all or some of the data in the shared ledger is encrypted.

Data is transmitted as \textit{transactions} from one party to another. In Bitcoin, for example, a transaction consists of some amount of bitcoin being sent from one address to another \cite{nakamoto08}.

The \textit{consensus algorithm} determines how each node adds blocks of new transactions to its copy of the shared ledger. Examples of consensus algorithms include Byzantine Fault Tolerance replication \cite{Vukolic16} and Proof-of-Work \cite{nakamoto08} (see Table \ref{table:blockchain-comparison} for more consensus algorithms). \\

\noindent The key feature of blockchain is that it enables trust without requiring a central authority, since the \emph{truth} is determined by an agreed-upon consensus algorithm. As long as the majority of a network is not collaborating to pollute the blockchain with incorrect data, its integrity is guaranteed \cite{nakamoto08}.

After Bitcoin's rapid rise in popularity in the past few years, many different blockchain implementations have emerged; Table \ref{table:blockchain-comparison} contains a comparison of key aspects of a selection of such blockchain implementations.



\section{Related Work}

In recent years, many companies have explored blockchain projects and prototypes for their respective industries. The container shipping industry is no exception: A.P. Moller–Maersk Group (``Maersk'' hereinafter) has launched a project in collaboration with EY (Ernst \& Young) and Microsoft Corporation to launch a marine insurance blockchain that would help reduce the market's inefficiencies. A prototype of the platform has been built on Microsoft Azure to ``make auditing aspects of a shipping supply chain easier, to improve the tamper-resistance and sharing of data in realtime, and to enable many different parties to settle upon the terms of premiums in a more timely fashion'' \cite{hackett2017}. The platform was to be deployed in January 2018, but no further news has emerged about it.
\section{\passport - A Decentralized System for Freight Declaration}
\label{section:protocol}


\begin{table}[!t]
  \renewcommand{\arraystretch}{1.3}
  \centering
  \caption{Fields stored in container and package claims}
  \label{table:claims}
  \begin{tabular}{ p{2cm} L{2.5cm} L{2.5cm} }
    \hline
    \textbf{Field} & \textbf{Container claims} & \textbf{Package claims}\\
    \hline
    Container ID & \multicolumn{2}{L{5cm}}{The ISO 6346 container identification number \cite{ISO6346}} \\
    Shipment ID & Concatination of the ship's IMO number \cite{imonumber} and date of departure in ISO 8601 format \cite{ISO8601} & (N/A)\\
    Package ID & (N/A) & This package's identification number, defined by the shipper\\
    From & Economic operator who has this container & (N/A)\\
    To & Economic operator who receives this container & (N/A)\\
    Sender & (N/A) & Person / company who sends this package\\
    Receiver & (N/A) & Person / company who receives this package\\
    Time & \multicolumn{2}{L{5.5cm}}{Time of claim submission in ISO \cite{ISO8601} format}\\
    Location & \multicolumn{2}{L{5.5cm}}{Longitude and latitude of the location at which the claim was made}\\
    Weight & Weight of this container, in kilograms & Weight of this package, in kilograms\\
    Action & (N/A) & \texttt{INSERT} if package was inserted into container; \texttt{REMOVE} if it was removed"\\
    Contents & (N/A) & Description of this package's contents\\
    \hline
  \end{tabular}
\end{table}

\passport{} is a secure and privacy-preserving decentralized system for freight declaration. It is built on a blockchain consisting of a network of certified nodes managed by \emph{customs agencies} and \emph{economic operators}. 
For \passport{}, we assume that:

\begin{enumerate}
	\item \emph{Economic operators trust the customs agency of their own country.} \passport{} must be implemented on a permissioned blockchain to ensure that only verified economic operators can submit data; the party that issues the certificates to enable this participation is the customs agency of the economic operator's country of origin; so they must be trusted to not abuse those certificates.
	\item \emph{Customs agencies do not trust customs agencies outside their trade bloc.} If there was complete trust between all customs agencies, a centralized system could be maintained by one of them; this is not the case.
    \item \emph{Packages in the system only move by shipping container.} We do not consider the movement of packages by other modes of transportation.
\end{enumerate}

Based upon these assumptions, we aim to achieve the following goals with \passport{}:


\subsubsection{Privacy-preserving}
In order to preserve the privacy of economic operators \passport{} must support visibility restrictions on the claims that economic operators post about the goods they are transporting. Most importantly, economic operators must not have the ability to read the data in each others' claims. Furthermore, only the customs agencies that are in the trade bloc that the package is destined for should be able to read the data in the claims, since economic operators do not necessarily trust every customs agency.

\subsubsection{Secure}
Non-repudiation is key to the security of the system because we want to make sure an economic operator can never deny a claim they made. To achieve this \passport{} must enforce that no economic operator in the system can alter previously submitted data. Economic operators can only submit new data when they have been granted access to the system by customs agencies. Claims about containers should only be accepted as long as the economic operators that submitted them are likely interacting with the containers.

\subsubsection{Scalable}
\passport{} must be able to handle enough transactions to support economic operators submitting transactions at any time. 
To support a gradual rollout, \passport{} must be able to track containers and packages even when not all economic operators in the supply chain participate in the system.

\subsubsection{Decentralized} \passport{} must be decentralized to avoid the shortcomings of a centralized system, such as the potential for a central database to be manipulated.


\subsection{Protocol Overview}
\label{section:protocol-overview}

\begin{algorithm*}
\caption{Validate container claim $C_X = X \xrightarrow{c} Y \mid X$ by operator $X$ about container $c$, with validation pool $P_c$.}\label{algorithm:validate-container-claim}
\label{algorithm:validation}
\begin{algorithmic}
\Procedure{Validate}{$C_X, P_c$} 
\State $T\gets\Call{query}{c} $\Comment{Query blockchain for latest accepted claim about $c$}
\If{\Call{isCustoms}{$X$}}
    \State \Call{accept}{$C_X$}
    		\Comment{Start a new trusted chain}
\ElsIf{$T \ \text{exists}$ \textbf{and} $T.\text{to} = C_\text{X}.\text{from}$}
    	\If{$C_\text{Y} \in P_c$ \textbf{and} $C_\text{X}\text{.from} = C_\text{Y}\text{.from}$ \textbf{and} $C_\text{X}\text{.to} = C_\text{Y}\text{.to}$}
        		\Comment{If matching claim by $Y$ exists in $P_c$}
        	\State \Call{accept}{$\{C_\text{X},C_\text{Y}\}$}
            		\Comment{Accept both claims}
            \State $P_c \gets \emptyset$ \Comment{Clear $P_c$ of wrong claims to save memory}
        \Else
        		\Comment{If matching claim by $Y$ does not (yet) exist in $P_c$}
    		\State $P_c \gets P_c \cup \{C_\text{X}\}$
            		\Comment{Add $C_\text{X}$ to $P_c$}
        \EndIf
\Else
    \State \Call{reject}{$C_X$}\Comment{Only customs agencies may create new trusted chains}
\EndIf
\EndProcedure
\end{algorithmic}
\end{algorithm*}

To meet these goals given our assumptions, we define \passport{} with the following entities:
\begin{itemize}
	\item \textit{Economic operators} are companies that either insert or remove packages from containers or transport containers. Economic operators submit data about the goods that they are handling to the blockchain network in the form of \textit{claims}.
	\item \textit{Customs agencies} process these claims and attempt to reach consensus over whether or not to append submitted claims to their shared ledger. 
	\item \textit{Containers} carry one or multiple packages of goods in them, and are identified by a container number that is specified according to the ISO 6346 standard \cite{ISO6346}. 
	\item \textit{Packages} are identified by the combination of container number, time and a number identifying them inside the container.
	\item Data that is submitted to the blockchain network about containers or packages are referred to as \textit{container claims} and \textit{package claims}. Claims are data objects that are signed by an economic operator.
\end{itemize}

The \passport{} protocol consists of the following three sub-protocols: 
\begin{enumerate}
	\item The \textit{claim submission protocol} (Section \ref{section:protocol-submission}) specifies how economic operators submit data to the blockchain. It is run on nodes belonging to economic operators.
    \item The \textit{container claim validation protocol} (Section \ref{section:protocol-validation}) determines whether data submitted by economic operators is valid and should be added to the shared ledger. It is run on nodes belonging to customs agencies.
    \item The \textit{economic operator certification protocol} (Section \ref{section:protocol-certification}) lets customs agencies allow or revoke access to economic operator in the blockchain network. The certification protocol is run on nodes belonging to customs agencies.
\end{enumerate}

We describe each sub-protocol in detail in the following sections.

\begin{figure}[!t]
	\small
	\centering
    \captionsetup[subfigure]{font=scriptsize,labelfont=scriptsize}
	\subfloat[\label{figure:chain-short}Trusted chain of transactions accepted by the validation algorithm.] {
        \makebox[.9\columnwidth]{\includegraphics[height=1.8cm]{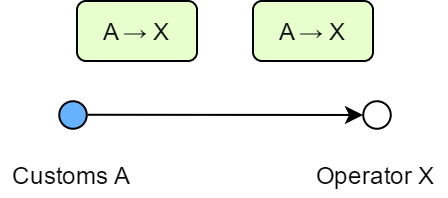}}
	} \\
	\subfloat[\label{figure:chain-long-pending}Chain of transactions with a new claim waiting for a match in the validation pool.] {
    	\makebox[.9\columnwidth]{\includegraphics[height=1.8cm]{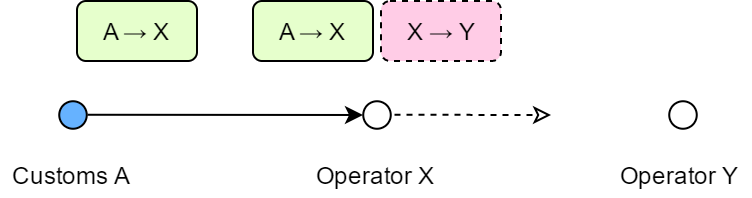}}
	} \\
	\subfloat[\label{figure:chain-long}Trusted chain of transactions accepted by the validation algorithm with a new transaction of two claims appended.] {
		\makebox[.9\columnwidth]{\includegraphics[height=1.8cm]{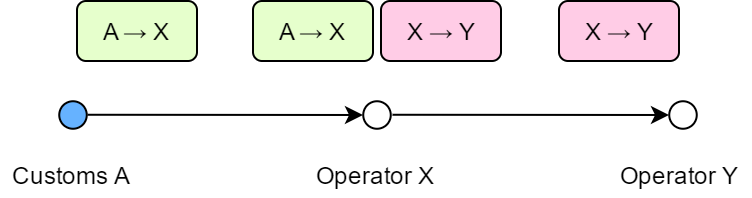}}
	}
\caption{\label{figure}Chains of transactions submitted by economic operators, same colors refer to the same transaction but made by different parties. a $\rightarrow$ b $\rightarrow$ c shows the process of adding a new transaction to the chain.} 
\end{figure}


\subsection{Claim Submission Protocol}
\label{section:protocol-submission}

Economic operators submit container claims and package claims to the network. Container claims tell customs agencies how containers are moving around the world, and package claims tell customs agencies what packages of goods are inside these containers. The data fields in each of these types of claims are described in Table \ref{table:claims}. 
We depict a \textit{container claim} as $X \xrightarrow{c} Y \mid S$, where $X$ is the economic operator that hands container $c$ to economic operator $Y$. $S$ is the signer of this claim and must be the same as either $X$ or $Y$. In the case that the next operator does not participate in the system, an economic operator $A$ should claim $A \xrightarrow{c} \epsilon \mid A$. Then, When an operator $B$ that participates in the system receives $c$ again, $B$ should claim $\epsilon \xrightarrow{c} B \mid B$. Before claims are appended to the shared ledger, customs agencies will run the validation protocol on claims.
Package claims must be encrypted using the public key of the destination's customs agency before they are submitted. Economic operators also add a plain-text field to package-claims to indicate which customs agency has the private key that can be used to decrypt the claim.


\subsection{Validation Protocol} 
\label{section:protocol-validation}

Algorithm \ref{algorithm:validate-container-claim} validates a submitted container claim and determines whether it should be added to the shared ledger. It first checks that the economic operator in the \textit{from} part of the new claim is actually in possession of the claim. It then makes sure that both economic operators involved in the claim agree on what happened. If both of these conditions are true, the claims are added to the shared ledger.

Given a container claim $C_X = X \xrightarrow{c} Y \mid X$, a previously accepted transaction of $A$ to $X$, and container $c$'s validation pool $P_c$, Algorithm \ref{algorithm:validation} first queries the blockchain for the latest claim $T$ about $c$, as shown in Figure \ref{figure:chain-short}. If $T$ exists and $c$ has been given to $X$ (which is the case in the example), then the new claim will be added to the pool of to-be-validated claims $P_c$. This scenario is shown in Figure \ref{figure:chain-long-pending}. The claim $X \xrightarrow{c} Y \mid X$ will only be accepted when $X \xrightarrow{c} Y \mid Y$ is submitted as shown in Figure \ref{figure:chain-long}.
Claims are not required to be submitted in the order of from-operator then to-operator, they are always added to $P_c$ when no matching claim is found.

If $c$ has not been given to $X$ in the previous transaction, the new claim can only be accepted if a customs agency has submitted the claim to reset the chain. This is required when operators can not confirm transactions and therefore the trusted chain is broken.

To save memory in the validation, all claims in the to-be-validated pool $P_c$ about container $c$ can be cleared when a claim is accepted for $c$. Also claims that contain impossible data can be left out of the validation pool.



\subsection{Certification Protocol}
\label{section:protocol-certification}

When an economic operator is to be added to the network, the customs agency in its country can certify that operator. This is done by generating a digital certificate that is signed by the newly added economic operator to prove that it has the correct key. When an economic operator misbehaves, the customs agency in their country can revoke the operator's certificate to restrict access to the blockchain. A revocation list is kept by the certificate authority, and blockchain nodes verify that claims have signatures that are generated using certificates that do not occur in the revocation list.

To add or remove customs agencies from the system, the customs agency nodes participate in a vote to reach consensus.

\section{System Design and Analysis}
\label{section:implementation}

\begin{figure}[!t]
\centering
\includegraphics[width=3.5in]{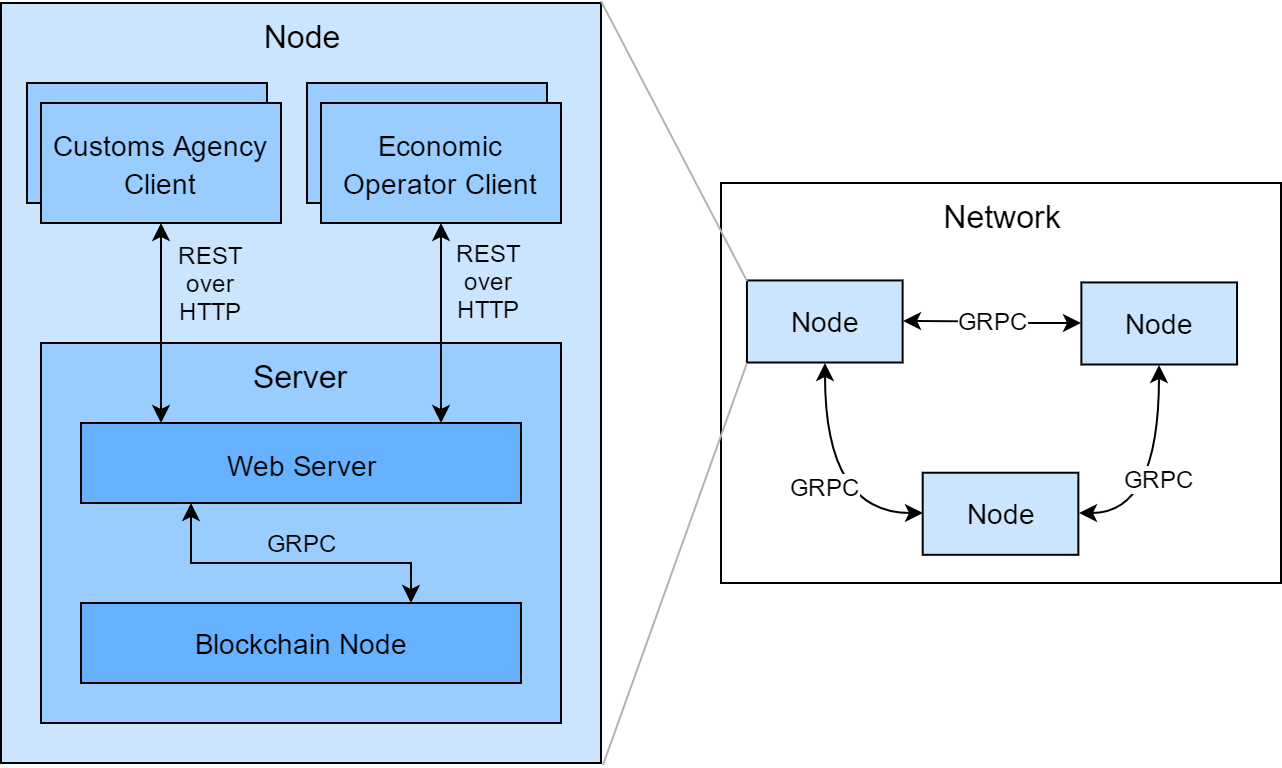}
\caption{System architecture for PassPort.}
\label{figure:architecture}
\end{figure}

We present a reference implementation for \passport{}, which implements all the protocols in Sections \ref{section:protocol-submission}, \ref{section:protocol-validation} and \ref{section:protocol-certification}. As shown in Figure \ref{figure:architecture}, several \emph{nodes} exist on the blockchain network. Such nodes may be \emph{customs agency nodes} (e.g. nodes run by customs agencies that follow the customs agencies protocol) or \emph{economic operator nodes} (e.g. nodes run by economic operators that follow the economic operator protocol). These nodes communicate with each other over gRPC\footnote{gRPC stands for ``gRPC Remote Process Call.'' See \url{http://www.grpc.io}.}. Each node has its own web server, which wraps its functionality in an API, with which it communicates over gRPC as well. The clients each communicate with the web server via REST over HTTP.


In the next sections, we describe each component in detail.

\subsection{System Design}
\subsubsection{Blockchain}
\label{section:implementation-blockchain}
We implement the blockchain component using Hyperledger Fabric \cite{cachin2016architecture}. We compare several blockchain implementations in Table \ref{table:blockchain-comparison}. Hyperledger Fabric is ideal for our protocol since it offers a private permissioned blockchain that supports at least 1,000 transactions per second (see Section \ref{section:system-analysis}) and allows for a pluggable custom consensus algorithm. 





\subsubsection{Client-Server Interaction}
\label{section:client-server-interaction}
For customs agencies and economic operators to interact with their peers in the blockchain network we implement a web server that wraps networking complexity in a RESTful API. The server receives API calls from two GUIs: One for economic operators and one for customs agencies. The customs agencies get an overview of shipments and containers, along with their estimated risk level. The GUI for economic operators mainly comprises of forms that are used to submit claims.


\subsection{Analysis}
\label{section:system-analysis}
In Section \ref{section:protocol} we have put forward some goals for \passport{} that we have addressed in our implementation. We now evaluate \passport{} in regards to these requirements.

\subsubsection{Privacy-preserving}
Since package data is encrypted using asymmetric encryption, only entities that have access to the private key can read the data. In the case of our protocol only the customs agency that will receive a package in their port has access to that private key. This means that economic operators can never read package data from others, only customs agencies that they will definitely interact with can.

\subsubsection{Secure}
To ensure that economic operators can only submit claims if they have been granted access to the blockchain network, customs agencies run a \emph{certificate authority node}, which grants the economic operators a key, which is part of their certificate, that they can use to sign their claims. Nodes in the system will immediately reject claims that are signed by revoked certificates.

We also introduce a protocol that ensures that economic operators can only make claims about containers they likely interacted with; economic operators' claims are in fact transactions. Economic operator $X$ can therefore only make a claim about container $c$ if there is another economic operator $Y$ that made a claim saying, $Y$ provided $X$ with $c$, and thus confirms $X$'s claim.

\subsubsection{Scalable}
Because economic operators must have the ability to make claims to other economic operators that are not part of the system we introduce an $\epsilon$ operator that represents a hole in the system. This means that containers can leave and reenter the system.

For a scalable system we must support sufficient throughput. We determine the amount of transactions that \passport{} must support as follows. As of 2012, there are approximately 32.9 million TEU\footnote{Twenty-foot Equivalent Unit, a standard shipping container size} shipping containers globally \cite{containerVolume} (order of magnitude: $10^8$). Because a single voyage takes days or weeks to complete, we generously estimate that a single container may `switch hands' up to $10^2$ times per year. Each time a container `switches hands', this requires a transaction. We estimate the amount of containers switching hands per second in Equation \ref{eq:containerVolume}:
\begin{equation}
	\frac{
    	10^8 \ \text{containers} 
    	\times 10^2 \ \tfrac{\text{moves}}{\text{year}} 
    }{
    	3600\times24\times365 \ \tfrac{\text{seconds}}{\text{year}}
    }
    \approx 317 \ \tfrac{\text{container moves}}{\text{second}}
    \label{eq:containerVolume}
\end{equation}

The only nodes that participate in the consensus algorithm are customs agencies, in the World Customs Organization 182 countries are included\cite{wco-members}. In practice, however, countries have formed customs unions reducing the need for every country to run their own node. Currently shipping is dominated by trade between 24 customs unions\cite{customs-unions}. Therefore the system must be able to support a maximum of approximately 24 nodes.

Our system must also consider what goods go into containers. If containers have a Full Container Load (FCL), this is one `package' per container per trip, but they have a Less than Container Load (LCL), this means multiple `packages' per container per trip. 

So \passport{} must have a throughput of at least $10^3$ transactions per second and be able to support around 24 nodes. Hyperledger Fabric has been shown to support this throughput in recent benchmarks and currently supports up to 16 nodes\cite{dinh2017blockbench}. With the release of Hyperledger Fabric version 1.1 a promise for higher scalability and performance have been made\cite{pc-improvements}.

\subsubsection{Decentralized}
Our implementation is decentralized as it uses a blockchain framework that runs on multiple nodes, it ensures that no single party controls the data in the system. By doing so, we remove trust that is required between parties.

\section{Conclusion}
\label{section:conclusion}

The shipping industry is responsible for the movement of millions of containers every day. Before these containers may enter a trade bloc, they must be cleared by the relevant customs agency. Because of the high volume of containers that must be processed each day, customs agencies perform risk analysis to decide which containers to audit. Risk analysis requires lots of data, which in the current system is potentially vulnerable to manipulation and malpractice because it is centralized and collected by a single authority.

In this work we have presented \passport{}, \textit{a secure and privacy-preserving decentralized system for freight declaration} that does not require the trust between entities that is required in centralized systems. In \passport{}, economic operators make \textit{claims} about the packages of goods and containers with which they interact, customs agencies validate those claims. Customs agencies and economic operators participate in a blockchain that validates this data and stores it in a secure and privacy-preserving manner. Our two key contributions are a data partitioning scheme and several protocols to enable this, and a reference implementation built on Hyperledger Fabric.

Firstly, our data partitioning scheme and protocols allow \passport{} to take advantage of the powerful validation principles enabled by blockchain, while hiding certain parts of the data to preserve the privacy of the involved economic operators. In our system, claims about the movement of containers are unencrypted, and can be validated to ensure that 1) the claim fits in the preceding chain of claims about that container and that 2) both parties involved in the claim agree on its contents. Claims about packages are encrypted so that only the customs agency at the goods' country of destination can see them. Hiding this critical link in the data means that only the appropriate customs agency can recreate the exact path that goods took to get to their country. This knowledge can improve the customs agency's risk analysis.

Secondly, our reference implementation built on Hyperledger Fabric shows that it is possible to implement \passport{} on a blockchain that meets our privacy-preservation, security, scalability, and decentralization requirements.

In future work, the combined container claim and package claim data provided by \passport{} could be used to further automate customs agencies' taxation procedures.

\bibliography{main.bib}
\bibliographystyle{IEEEtran}

\end{document}